\documentclass[aps,prl,amsmath,twocolumn]
{revtex4}
\usepackage[dvips]{graphicx}
\newcommand{\PSbox}[3]{\mbox{\rule{0in}{#3}\includegraphics{#1}\hspace{#2}}}

\begin{document}
\draft

\title{
The Cyclotron Spin-Flip Mode as the Lowest-Energy Excitation
 of Unpolarized Integer Quantum Hall States}  \vskip -3.mm
\author{L.V.~Kulik,$^{a,b}$ I.V.~Kukushkin,$^{a,b}$ S.~Dickmann,$^{a}$ V.E.~Kirpichev,$^{a,b}$ A.B.~Van'kov,$^{a}$
A.L.~Parakhonsky,$^{a}$ J.H.~Smet,$^{b}$ K.~von Klitzing,$^{b}$ W.
Wegscheider$^{c,d}$}

\affiliation{$^a$Institute for Solid State Physics, Russian
Academy of Sciences, Chernogolovka, 142432 Russia\\
$^b$Max-Planck-Institut f\"ur Festk\"orperforschung,
Heisenbergstra\ss e 1, 70569 Stuttgart, Germany\\ $^c$Walter
Schottky Institut, Technische Universit\"at M\"unchen, 85748
Garching, Germany\\
$^d$Institut f\"ur Experimentelle und Angewandte Physik,
Universit\"at Regensburg, 93040 Regensburg, Germany }

\begin{abstract}

The cyclotron spin-flip modes of spin unpolarized integer quantum
Hall states ($\nu =2,4,6$) have been studied with inelastic light
scattering. The energy of these modes is significantly smaller
compared to the bare cyclotron gap. Second order exchange
corrections are held responsible for a negative energy
contribution and render these modes the lowest energy excitations
of unpolarized integer quantum Hall states.\vspace{2.mm}


\end{abstract}

\maketitle

According to Kohn's theorem \cite{ko61}, homogenous
electromagnetic radiation incident on a translationally invariant
electron system can only couple to the center-of-mass coordinate.
Such radiation is unable to excite internal degrees of freedom
associated with the Coulomb interaction. As a result, physical
phenomena originating from electron-electron interactions leave
the cyclotron resonance unaffected. Hence, spin-unperturbed
magnetoplasmons excited under these conditions have an energy
equal to the bare cyclotron energy, irrespective of existing
electron-electron correlations \cite{kallin84}. A similar
statement also holds for spin-excitons, intra-Landau level
spin-flip excitations. In a system with rotational invariance in
spin space, Larmor's theorem \cite{dobbers88} dictates that
Coulomb interactions do not contribute to the energy of
zero-momentum spin-excitons. In contrast to these magnetoplasma
and spin-exciton excitations, there exist no symmetry arguments
which restrict the energy of the combined zero momentum cyclotron
spin-flip mode (CSFM). It is well established that the
cyclotron-spin-flip mode excited from spin-polarized ground states
acquires considerable exchange energy even for zero
momentum~\cite{pinczuk92,kallin93,kulik01}. The energy of this
mode may thus serve as a unique probe of many-body interactions in
the electronic system.

Hitherto, it has not been considered that there is also an
exchange contribution to the energy of the zero momentum
cyclotron-spin-flip modes of {\it unpolarized} quantum Hall ground
states at even integer fillings ($\nu = 2, 4, 6, \ldots$). First
order perturbation calculations in the ratio $r_{\rm c}\!=\!E_{\rm
C}\!/\hbar\omega_{\rm c}$ explicitly predicted a zero exchange
contribution to the total energy of this combined mode of
unpolarized quantum Hall ground states~\cite{kallin84} ($E_{\rm
C}$ is the characteristic Coulomb energy scale and
$\hbar\omega_{\rm c}$ the cyclotron energy). Here, we
experimentally demonstrate however that the energy of these modes
is considerably reduced compared with the bare cyclotron gap. We
corroborate with theoretical considerations that the negative
energy contribution arises from {\it second order Coulomb
corrections} and so was not captured by previous first order
perturbation calculations.

Two high-quality heterostructures were studied. Each consisted of
a single-side modulation doped $30\ {\rm nm}$
 AlGaAs/GaAs quantum well (QW) with an electron density
between $1$ and $1.2 \cdot 10^{11}$~cm$^{-2}$ and a mobility of
$5-7 \cdot 10^6$~cm$^2$ /(Vs). The density $n_s$ was tuned
continuously via the opto-depletion effect and was measured with
luminescence \cite{kukushkin}. Inelastic light scattering (ILS)
spectra were recorded at $1.5\ {\rm K}$ in the back-scattering
geometry in a split-coil cryostat. Three optical fibers were
utilized. One fiber transmitted a dye laser pump beam, tuned above
the fundamental gap of the QW. The remaining fibers collected the
scattered light and guided it out of the cryostat. The angles
between the sample surface, pump beam fiber and collecting fibers
define the in-plane momentum transferred to the electron system
via inelastic light scattering. The collecting fibers selected
excitations with in-plane momenta of $0.4$ and $1.0 \cdot
10^5$~cm$^{-1}$. The scattered light was dispersed in a triple
grating monochromator and detected with a CCD camera.

\begin{figure}[tb!]
\PSbox{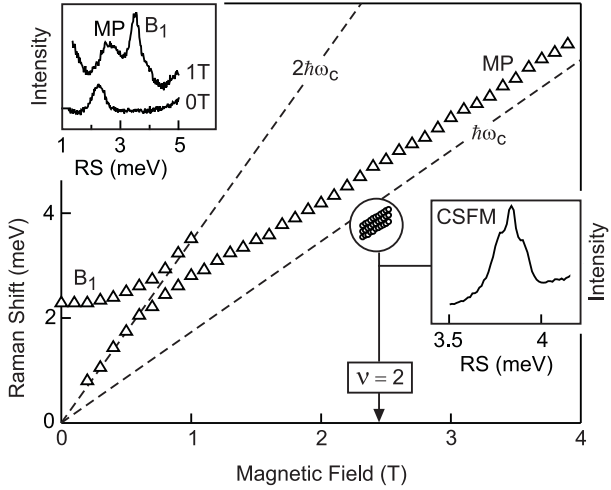}{8cm}{7cm} \caption{$B$-dependence of ILS line
energies. Triangles mark charge density excitations, circles the
CSFM. The dashed lines correspond to $\hbar \omega_{\rm c}$ and $2
\hbar \omega_{\rm c}$. Top, left inset: ILS spectra at $B\!=\!0$
and $ 1\ {\rm T}$. Bottom, right inset: ILS spectrum at $2.4\ {\rm
T}$.}\label{fig1}\vspace{-3.mm}
\end{figure}

Fig.~\ref{fig1} shows typical ILS spectra of inter Landau-level
(LL) excitations as well as the magnetic field ($B$) dependence of
the energy of the various lines in these spectra in a sample with
a density of $1.2 \cdot 10^{11}\,$cm$^{-2}$. The experimental
configuration selected excitations with an in-plane momentum of $q
= 1.0 \cdot 10^5\,$cm$^{-1}$. The polarization selection rules
allowed to identify that lines at low $B$ ($<\!1\,$T) correspond
to charge density excitations. The principal magnetoplasmon  mode
as well as a Bernstein mode (B$_1$) are observed in the geometry
where the incident and scattered photons have parallel
polarization vectors \cite{abstreiter84}. At non-zero $B$ the
magnetoplamon mode has a strong linear dispersion in the long
wavelength limit and at $B\!=\!0$ its energy equals the plasma
energy for momentum $q$. In contrast, the Bernstein mode is nearly
dispersionless. Both modes couple through many body Coulomb
interactions near $\sim\!0.8\,$T. At large $B$, their energies
converge asymptotically to the cyclotron energy and twice the
cyclotron energy, respectively \cite{batke}.

Of main interest here is the appearance of a triplet ILS resonance
when the system is in the $\nu = 2$ spin-unpolarized quantum Hall
state (bottom, right inset Fig.~\ref{fig1}). Near $B = 2.4\ {\rm
T}$, the central line of the triplet is clearly resolved, but the
side lines only appear as shoulders. The splitting between the
features corresponds approximately to the electron Zeeman energy
$E_{\rm Z}$ in GaAs, so they are attributed to the three cyclotron
spin-flip modes with different spin projections along the
$B$-field axis ($S_z=-1,0$ and $1$). The shoulder structures are
assigned to the cyclotron spin-flip modes with $S_z\!=\!-1$ and
$1$, and the central line ($S_z\!=\!0$) is associated with a
cyclotron spin-wave, i.e.~out-of-phase oscillations of the two
spin subsystems of the Landau levels with orbital index 0 and 1
\cite{kulik02}. This identification of the triplet is confirmed by
measurements in tilted fields. ILS spectra in tilted fields are
plotted in the left inset to Fig.~2. The triplets are much better
resolved due to the larger total fields $B_{\rm tot}$.
Well-separated peaks appear and the spin splitting can be directly
measured. The Zeeman effect is in essence a three dimensional
phenomenon and so energy gaps between the ILS triplet lines are
proportional to $B_{\rm tot}$ rather than the perpendicular
component $B_{\perp}$. The position of the central line however
only depends on $B_{\perp}$ (left inset Fig.~2). The main plot in
Fig.~2 presents the measured electron Zeeman energy (open circles)
as a function of $B_{\rm tot}$. The data points fit well to a
$g$-factor $g_{\mbox{\tiny QW}}=-0.4$ (solid line). The dashed
line corresponds to $|g_{\mbox{\tiny GaAs}} \mu_{\rm B} B|$, where
$g_{\mbox{\tiny GaAs}}=-0.44$ is the effective g-factor of bulk
GaAs. A significant reduction of the g-factor is not uncommon in
AlGaAs-heterostructures and has been accounted for by
bandstructure non-parabolicity, confinement and wavefunction
penetration effects \cite{rossler}.

In the inset of Fig.~3 we compare ILS spectra measured at $\nu=2$
for two different values of in-plane momenta: $0.4$ and $1.0
\times 10^5$~cm$^{-1}$. In agreement with
 existing theories~\cite{kallin84, mac86}, the CSFM energy does
not show any appreciable dispersion at momentum values accessible
with ILS techniques. Therefore, the CSFM line is regarded as the
energy of the cyclotron spin-flip mode when $q \rightarrow 0$. The
key experimental finding is a downward shift of the energy of this
mode with $-0.35\ {\rm meV}$ as compared to the bare cyclotron
energy. This shift exceeds by far the single electron Zeeman
energy in GaAs at this magnetic field ($0.08\ {\rm meV}$) and we
therefore assert it is strongly influenced by exchange
interactions.

The $B$-dependence of the energy of the cyclotron spin flip mode
for fixed filling $\nu =2$ is plotted in Fig.~3. The slope is
identical to the bare cyclotron energy line in GaAs. Hence, the
dependence of the cyclotron spin flip mode on $B$ takes on the
functional form $\hbar \omega_{\mbox{\tiny
   CSFM}} = \hbar \omega_{\rm c} +
\Delta E_{\mbox{\scriptsize
   SF}}$ over a rather broad magnetic field
interval: $0.6\,T\! < \!B \!<\! 2.7\,T$. Here $\Delta
E_{\mbox{\scriptsize
   SF}}$ is the $B$-independent downward shift of approximately $-0.35\ {\rm meV}$. It
is worthwhile to note that a dimensional analysis of second order
Coulomb corrections to the energies of inter-LL excitations would
yield a similar dependence on $B$: $E\!=\!\hbar\omega_{\rm
c}\!+\!\Delta E_{\mbox{\scriptsize
   SF}}$, where $\Delta
E_{\mbox{\scriptsize
   SF}}\!\sim\!\hbar\omega_{\rm c} r_{\rm c}^2$. Indeed, if $E_{\rm
C}\!=\!\alpha e^2/\varepsilon l_B$ then $\Delta
E_{\mbox{\scriptsize
   SF}}$ is
independent of the field. The renormalization factor $\alpha$ is
determined by the size-quantized wave function of electrons
confined in the QW. In the ideal 2D case $\alpha\!=\!1$. However,
the larger the width of the 2D electron system (2DES), the smaller
$\alpha$ becomes and thereby reducing $r_{\rm c}$. This is
certainly relevant for the width of our quantum well.

An {\it analytical} calculation of the second order correction to
the CSFM energy is performed in terms of small $r_{\rm c}$. The
theory is based on the following general features of the system.
The state of the system is described by the exact quantum numbers
$S$, $S_z$ and ${\bf q}$ and by the 'good' quantum number $\delta
n$ characterizing the excitation kinetic energy $\hbar\omega_{\rm
c}\delta n$ ($\delta n$ is good but not exact due to LL-mixing).
The relevant excitations with ${\bf q}\!=\!0$ and $\delta n\!=\!1$
may be presented in the form ${\hat{K}}_{S\!,S_z}^{\dag}|{\bf
0}\rangle$, where $|{\bf 0}\rangle$ is the ground state and
${\hat{K}}_{S\!,S_z}^{\dag}$ are ``raising" operators:
$\!{\hat{K}}_{0\!,0}^{\dag}\!\!=\!\!\sum_{np\sigma}\!{}\!\sqrt{n+1}
c^{\dag}_{n\!+\!1\!,p,\sigma}c_{n,p,\sigma},
\;{\hat{K}}_{1\!,0}^{\dag}\!\!=\!\!\sum_{np\sigma}\!{}\!\sqrt{n+1}\,
(-1)^{\sigma}c^{\dag}_{n\!+\!1,p,\sigma}c_{n,p,\sigma}
\qquad\mbox{and}\qquad{\hat{K}}_{1\!,+/-}^{\dag}=\sum_{np}\!{}\!\sqrt{n+1}
c^{\dag}_{n\!+\!1\!,p,\uparrow/\downarrow}c_{n,p,\downarrow/\uparrow}$,
[$c_{n,p,\sigma}$ is the Fermi annihilation operator corresponding
to the Landau-gauge state $(n,p)$ with spin index
$\sigma\!=\!\uparrow,\downarrow$]. The commutators with the
kinetic-energy operator ${\hat{H}}_1$  are
$[{\hat{H}}_1,{\hat{K}}_{S\!,S_z}^{\dag}]\!\!\equiv
\!\!\hbar\omega_{\rm c}{\hat{K}}_{S\!,S_z}^{\dag}$. The total
Hamiltonian is $ {\hat H}_{\mbox{\scriptsize tot}}\!=\!{\hat
H}_1\!+\!{\hat H}_{\mbox{\scriptsize int}}$, where ${\hat
H}_{\mbox{\scriptsize int}}$ is the exact Coulomb-interaction
Hamiltonian. If $|{\bf 0}\rangle$ is unpolarized, we have
${\hat{\bf S}}^2{\hat{K}}_{S\!,S_z}^{\dag}|{\bf
0}\rangle\!\!\equiv\!\!S(S\!+\!1){\hat{K}}_{S\!,S_z}^{\dag}|{\bf
0}\rangle$ and ${\hat{S}}_z{\hat{K}}_{S\!,S_z}^{\dag}|{\bf
0}\rangle\!\!\equiv\!\!S_z{\hat{K}}_{S\!,S_z}^{\dag}|{\bf
0}\rangle$. Moreover, the identity $\langle {
0}|{\hat{K}}_{S\!,S_z}[{\hat H}_{\mbox{\scriptsize int}},
{\hat{K}}_{S\!,S_z}^{\dag}]|{ 0}\rangle\!\equiv\!0$ holds
($|0\rangle$, to describe the zero'th order ground state).
 It implies that first-order Coulomb
corrections vanish for both the spin unperturbed or singlet
magnetoplasmon (where $S\!=\!0$) and the combined CSFM triplet
($S\!=\!1$). At the same time, $[{\hat H}_{\mbox{\scriptsize
int}},{\hat{K}}_{0,0}^{\dag}]\!\equiv\!0$ \cite{ko61} but $[{\hat
H}_{\mbox{\scriptsize int}},{\hat{K}}_{1,S_z}^{\dag}]\!\neq\!0$.
Hence, whereas the magnetoplasmon has no exchange energy
correction calculated to {\it any order} in $r_{\rm c}$, the
combined modes should have second and higher order exchange
corrections.

The second-order calculation is based on the Excitonic
Representation (ER) technique \cite{dz83,di02,di96}. It utilizes
exciton states $\hat{\cal Q}_{ab\,{\bf q}}^{\dag}|0\rangle$ as a
basis set, instead of single-electron states of a degenerated LL.
The exciton creation operator is defined as \cite{dz83,di02,di96}
$$
 \hat{\cal Q}_{ab\,{\bf
 q}}^{\dag}\!=\!N_{\phi}^{-1/2}\mbox{$\sum_{p}$}
  e^{-iq_x\!p}\,
  b_{p\!+\!{q_y}/{2}}^{\dag}a_{p\!-\!{q_y}/{2}}.
   \vspace{-.5mm}\eqno (1)
$$
Here,  ${}\!N_{\phi}\!\!=\!\!A/2\pi l_B^2\!$ stands for the number
of magnetic flux quanta and ${}\!{\bf q}\!\!=\!\!(q_x,q_y)$ is
given in  units of ${}1/l_B\!{}$. The binary indices $a$ and $b$
denote both the LL number and the spin index:
$\!a,b\!\!=\!\!(n_{a,b},\sigma_{a,b})$. All three CSFM states have
certainly the same exchange energy, and it is sufficient to
calculate this, e.g., for the state with $S_z\!=\!-\!1$. The
zero-order approximation is thereby
$|SF,-\rangle\!\!=\!\!N_{\phi}^{-1/2}{\hat{K}}_{1,-}^{\dag}|{\bf
0}\rangle|_{r_{\rm c}\!=\!0}\!\!=\!\!\hat{\cal
Q}_{0\overline{1}\,{\bf 0}}^{\dag}|0\rangle$ [i.e.
$a\!=\!(0,\uparrow)$ and $b\!=\!(1,\downarrow)$]. To calculate the
first-order corrections to the $|SF,-\rangle$ state or,
equivalently, the second-order correction to its energy, we follow
the standard perturbative approach \cite{ll91} using the
``excitonically non-diagonilized" part ${\hat{\cal
H}}_{\mbox{\scriptsize {int}}}$ of the Coulomb Hamiltonian
\cite{di02} in the ER form: \vspace{-2.mm}
$$
  \begin{array}{l}
  {}\!{}\!{}\!{}\!{\hat{\cal  H}}_{\mbox{\scriptsize {int}}}=
  {\displaystyle \frac{e^2}{2\varepsilon l_B}}\sum_{{\bf q},a,b,c,d}
  V({\bf q})\left[h_{n_an_b}({\bf q})\delta_{\sigma_a,\sigma_b}
  {\hat {\cal Q}}_{ab{\bf
  q}}^{\dag}\right]\\{}\quad{}\qquad{}\qquad{}\qquad{}
  \qquad\times\!\left[h_{n_cn_d}(-{\bf q}) \delta_{\sigma_c,\sigma_d}
  {\hat {\cal Q}}_{cd\,-\!{\bf q}}^{\dag}\right]
  \end{array}  \vspace{-1mm}\eqno (2)
$$
(c.f. \cite{di02}), where $2\pi V({\bf q})$ is the Fourier
component of the dimensionless Coulomb potential (in the strict 2D
limit $V\!=\!1/q$), and $h_{kn}({\bf
q})=({k!}/{n!})^{1/2}e^{-q^2/4}
  (q_-)^{n\!-\!k}L^{n\!-\!k}_{k}(q^2/2)$ are the ER ``building-block''
functions ($L_k^n$ is the Laguerre polynomial,
$q_{\pm}=\mp\frac{i}{\sqrt{2}}(q_x\pm iq_y)$;
 c.f. Refs. \cite{kallin84,di02,di96}). For calculation details,
we refer the reader to Ref. \cite{c-m04}. Here we limit ourselves
to reporting the final result\vspace{-1.mm}
$$
   \Delta E_{\mbox{\scriptsize
   SF}}=-\sum\limits_{n=2}^{\infty}R_n\frac{1-2^{1-n}}{n(n^2-1)},
   \eqno (3)
$$
\vspace{-2.mm} with \vspace{-3.mm}
$$
 R_n\!=\!\frac{2}{n!}\!\int_0^{\infty}dqq^{2n\!+\!3}
V^2(q)e^{-q^2},
$$
in units of 2Ry$=\!(e^2/\varepsilon l_B)^2/\hbar\omega_{\rm
c}\!\approx\!11.34\,$meV. For the ideal 2D system with zero width
$R_n\!\equiv\!1$ and the summation can easily be performed. It
yields $\Delta E_{\mbox{\scriptsize SF}}=(\ln{2}-1)/2=-0.1534...$.

We conclude that, as in experiment, the exchange interaction
lowers the energy of the CSFM relative to the singlet
magnetoplasmon mode. The absolute value of the shift $|\Delta
E_{\mbox{\scriptsize
   SF}}|$ obtained with Eq.~3 is reduced when taking into
account the non-zero thickness of the 2DES. The Coulomb vertex
should be written as $V(q)\!=\!F(qw)/q$, where $F(qw)$ is a form
factor capturing the Coulomb softening~\cite{an82}. The effective
thickness parameter $w$ characterizes the spread of the electron
wavefunction in the growth direction. If a variational
wavefunction of the form $|\psi(z)|^2\!\sim\! \exp{(-z^2/2w^2)}$
is chosen, then $F(qw)\!=\!e^{w^2q^2}\mbox{erfc}(wq)~$\cite{co97}.
Note that for a second-order energy correction this form-factor
enters twice. The calculation of $|\Delta E_{\mbox{\scriptsize
   SF}}|$, including the influence of finite
thickness, is plotted in Fig.~4. A similar value for $|\Delta
E_{\mbox{\scriptsize
   SF}}|$ as in experiment is obtained when $w\approx 0.5l_B$, which
agrees well with the effective width for a $30\,$nm GaAs
QW-structure.

We note that inelastic light scattering studies at $\nu$=2 were
carried out previously in Ref.~\cite{Eriksson99}. The authors
obtained similar spectra with a non-zero energy shift, but
explained their observations in terms of transitions to the roton
minimum at large wave-vector. This assignment was plausible as it
was based on information from Ref.~[2] in which a non-zero energy
shift was predicted only for the roton minimum but not for $q=0$
due to the first order approximation in the interaction. The
authors were forced to invoke disorder to account for the large
momentum transfer required for scattering into the roton minimum.
The energy shift was reported to fit to a square root dependence
over the investigated $B$-field range as anticipated for rotons.
In contrast, we assign the spectra in our experiments to the
properties of the cyclotron spin-flip mode at $q=0$. If we were to
ascribe the signals to indirect inelastic light scattering into
the roton mimimum, we would expect a second much larger direct
resonance at $q=0$ since the density of states is large for both
the $q=0$ extremum and the roton minimum. Moreover, we find a
field independent energy shift $\Delta E_{SF}$ over a large
$B$-range. Well resolved triplet spectra and negative exchange
energy shifts are not only obtained at $\nu$=2, but also at
filling $\nu$=4 and $\nu$=6. Fig.~5 illustrates for instance
ILS-spectra measured at $\nu=4$. Note that the negative exchange
energy contribution at $\nu=4$ is only half of the value at
$\nu\!=\!2$ due to the larger spatial extent of the wave functions
of exciton states of higher Landau levels. The observation of well
resolved triplet modes at $\nu$=4 and $\nu$=6 excludes an
interpretation of our data at momenta of the roton minimum. In
this case, two and three roton minima would appear for $\nu = 4$
and $\nu = 6$. This would result in significant broadening and the
triplet structure would be smeared out.

In conclusion, the inelastic light scattering response from the
combined cyclotron spin-flip modes of unpolarized quantum Hall
states at $q=0$ has been studied. A negative energy term was found
to decrease their energy and was attributed to many-body Coulomb
exchange interaction. A second order perturbation theory of the
Coulomb interaction explains the experimental results.

\vspace{2.mm}

Financial support from Max-Planck and Humboldt Research Grants,
the Russian Fund of Basic Research,  INTAS (Project No. 03-
51-6453) and the BMBF is acknowledged.

\begin{figure}[tb!]
\PSbox{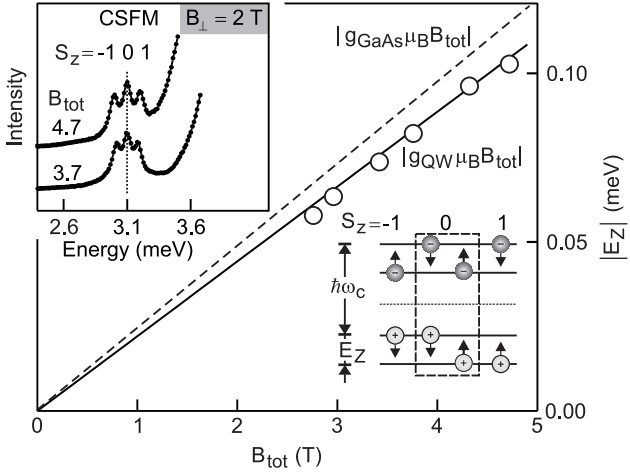}{8cm}{6.3cm} \caption{Electron Zeeman energy
$E_{\rm Z}$. The dashed line plots the expected Zeeman energy when
taking $g_{\mbox{\tiny GaAs}}\!=\! -0.44$. The solid line is a
linear fit to the data for $g_{\mbox{\tiny QW}}\!=\! -0.4$. The
inset depicts ILS spectra at a constant perpendicular field of 2T
but two different values of the total field. The right inset
schematically illustrates the spin-triplet
($S\!=\!1,\;\,S_{z}=-1,\,0$ and $ 1$) cyclotron excitations for
$\nu = 2$.}\label{fig2}\vspace{-4.mm}
\end{figure}

\begin{figure}[tb!]
\PSbox{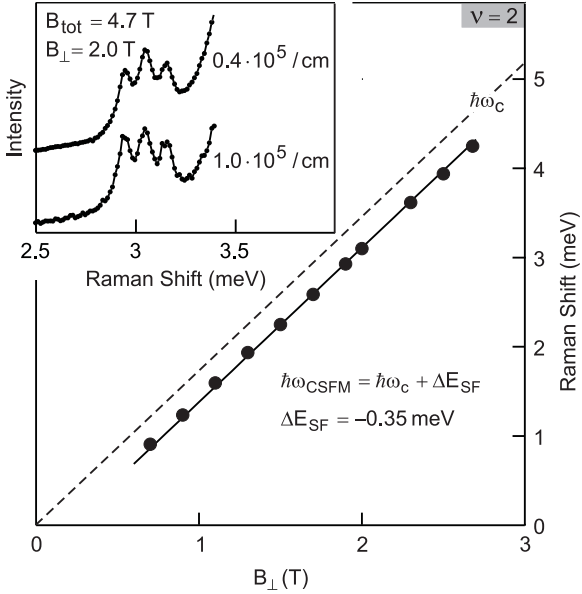}{8.0cm}{7.3cm} \vspace{-0.mm} \caption{The CSFM
energy for the spin unpolarized $\nu = 2$ quantum Hall state
versus perpendicular field. The dashed line gives the cyclotron
energy. The upper inset displays ILS spectra of the CSFM for two
different in-plane momenta.} \label{fig3}\vspace{-5.mm}
\end{figure}

\begin{figure}[tb!]
\PSbox{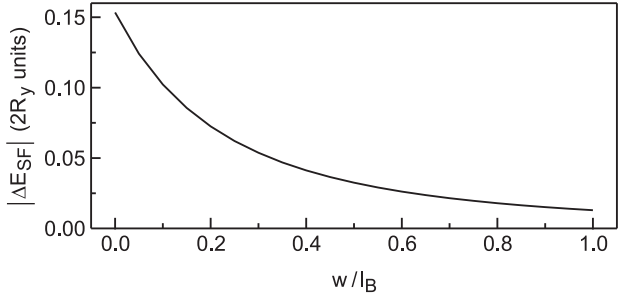}{8cm}{3.8cm} \vspace{-3.mm} \caption{The CSFM
exchange shift calculated from Eq. (3) with the modified Coulomb
interaction $V(q)\!=\!q^{-1}e^{q^2w^2}\mbox{erfc}(qw)$. Its
absolute value at $w\!=\!0$ equals
$(1-\ln{2})\cdot$Ry.}\vspace{-1.mm}
\end{figure}

\begin{figure}[tb]
\vspace{2.mm} \PSbox{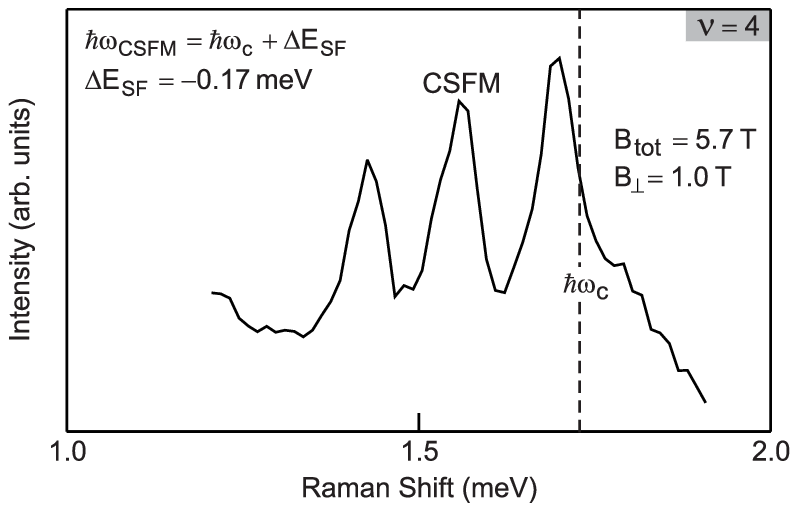}{8cm}{5.3cm}\vspace{1.mm}
\caption{ILS spectrum of the cyclotron spin-flip mode at the
$\nu\!=\!4$ unpolarized quantum Hall state at the indicated values
for $B_{\rm \perp}$ and $B_{\rm tot}$.} \label{fig5} \vspace{-5mm}
\end{figure}


\vspace{-2.mm}

\begin{references}

\bibitem{ko61}
W.~Kohn, Phys.~Rev.~{\bf 123}, 1242 (1961).

\bibitem{kallin84}
C.~Kallin and B.I.~Halperin, Phys.~Rev.\! B \!{\bf 30}, \!5655
(1984).

\bibitem{dobbers88}
M.~Dobers, K.~von Klitzing, G.~Weimann, Phys.~Rev.~B {\bf 38},
5453 (1988).

\bibitem{pinczuk92}
A.~Pinczuk, B.~S.~Dennis, D.~Heiman, C.~Kallin, L.~Brey,
C.~Tejedor, S.~Schmitt-Rink, L.~N.~Pfeiffer, K.~W.~West,
Phys.~Rev.~Lett.~{\bf 68}, 3623 (1992).



\bibitem{kallin93}
J.~P.~Longo and C.~Kallin, Phys.~Rev.~B {\bf 47}, 4429 (1993).

\bibitem{kulik01}
L.~V.~Kulik, I.~V.~Kukushkin, V.~E.~Kirpichev, J.~H.~Smet, K.~von
Klitzing, and W.~Wegscheider, Phys.~Rev.~B {\bf 63}, R201402
(2001).

\bibitem{kukushkin}
I.~V.~Kukushkin and V.~B.~Timofeev, Adv.~Phys.~{\bf 45}, 147
(1996).

\bibitem{abstreiter84}
G. Abstreiter, M. Cardona, and A. Pinczuk, in {\it Light
Scattering in Solid IV} edited by M. Cardona and G. Guntherodt
(Springer-Verlag, Berlin, 1984), p.5.

\bibitem{batke}
E.~Batke, D.~Heitmann, J.~P.~Kotthaus, and K.~Ploog,
Phys.~Rev.~Lett. {\bf 54}, 2367 (1985).

\bibitem{kulik02}
L.~V.~Kulik, I.~V.~Kukushkin, V.~E.~Kirpichev, J.~H.~Smet, K.~von
Klitzing, V.~Umansky, and W.~Wegscheider, Physica E {\bf 12}, 574
(2002).

\bibitem{rossler}
G.~Lommer, F.~Malcher, and U.~Rossler, Phys.~Rev.~B {\bf 32}, 6965
(1985).

\bibitem{mac86}
H.~C.~A.~Oji and A.~H.~MacDonald, Phys.~Rev.~B {\bf 33}, 3810
(1986).

\bibitem{dz83}
{A.~B.~Dzyubenko and Yu.~E.~Lozovik}, Sov.~Phys.~Solid State {\bf
25}, 874 (1983) [{\it ibid}. {\bf 26}, 938 (1984)].

\bibitem{di02}
S.~Dickmann, Phys. Rev. B {\bf 65}, 195310 (2002).

\bibitem{di96}
S.~Dickmann, S.~V. Iordanskii, JETP {\bf 83}, 128 (1996);
S.~Dickmann, Y.~Levinson, Phys.~Rev.~B {\bf 60} 7760 (1999);
S.~Dickmann, Phys.~Rev.~B {\bf 61}, 5461 (2000).

\bibitem{ll91}
L.D. Landau and E.M. Lifschitz, {\it Quantum Mechanics}
(Butterworth-Heinemann, Oxford, 1991).

\bibitem{c-m04}
S.~Dickmann and I.~V.~Kukushkin, cond-mat/0411707.

\bibitem{an82} T.~Ando, A.~B.~Fowler, F.~Stern, Rev.~Mod.~Phys.~{\bf 54},
437 (1982).

\bibitem{co97}
N.~R.~Cooper, Phys.~Rev.~B {\bf 55}, 1934 (1997).

\bibitem{Eriksson99}
M.~A.~Eriksson, A.~Pinczuk, B.~S.~Dennis, S.~H.~Simon,
L.~N.~Pfeiffer, K.~W.~West, Phys.~Rev.~Lett.~{\bf 82}, 2163
(1999).

\end{references}
\end{document}